\documentclass[superscriptaddress,prb,twocolumn,aps,nobibnotes]{revtex4-2}

\usepackage{graphicx}
\usepackage{xcolor}
\usepackage{multirow}
\usepackage{booktabs}
\usepackage{array,mathtools}
\usepackage{xspace}
\usepackage{tikz}

\newcommand{\mna}{Mn$_{4a}$\xspace}
\newcommand{\mnb}{Mn$_{4b}$\xspace}
\newcommand{\mnc}{Mn$_{8c}$\xspace}
\newcommand{\Sn}{Ni$_{2}$Mn$_{1.5}$Sn$_{0.5}$\xspace}
\newcommand{\In}{Ni$_{2}$Mn$_{1.5}$In$_{0.5}$\xspace}
\newcommand{\InSn}{Ni$_{2}$Mn$_{1.5}$(Sn$_{x}$In$_{1-x}$)$_{0.5}$\xspace}
\newcommand{\alloys}{Ni$_{1.92}$Mn$_{1.44}$(Sn$_x$In$_{1-x}$)$_{0.64}$\xspace}
\newcommand{\alloysn}{Ni$_{1.92}$Mn$_{1.44}$Sn$_{0.64}$\xspace}
\newcommand{\alloyin}{Ni$_{1.92}$Mn$_{1.44}$In$_{0.64}$\xspace}
\begin{document}

\title[]{Magnetic phase diagram of the austenitic Mn-rich Ni-Mn-(In,Sn) Heusler alloys}

%
%
\author{P. Bonfà}
\affiliation{Department of Mathematical, Physical and Computer Sciences, University of Parma, 43124 Parma, Italy}
\affiliation{Centro S3, CNR-Istituto Nanoscienze, 41125 Modena, Italy}

\author{S. Chicco}
\affiliation{Department of Mathematical, Physical and Computer Sciences, University of Parma, 43124 Parma, Italy}
\affiliation{UdR Parma, INSTM, I-43124 Parma, Italy}

\author{F. Cugini}
\affiliation{Department of Mathematical, Physical and Computer Sciences, University of Parma, 43124 Parma, Italy}
\affiliation{Institute of Materials for Electronics and Magnetism, National Research Council (IMEM-CNR), Parco Area delle Scienze 37/A, I-43124 Parma, Italy}

\author{S. Sharma}
\affiliation{Max-Born-Institute for Non-linear Optics and Short Pulse Spectroscopy, Max-Born Strasse 2A, D-12489 Berlin, Germany}

\author{J.~K. Dewhurst}
\affiliation{Max-Planck-Institut fur Mikrostrukturphysik Weinberg 2, D-06120 Halle, Germany}

\author{G. Allodi}
\affiliation{Department of Mathematical, Physical and Computer Sciences, University of Parma, 43124 Parma, Italy}







\begin{abstract}
Heusler compounds have been intensively studied owing to the
important technological advancements that they provide in the field of shape memory, thermomagnetic energy conversion
and spintronics.
Many of their intriguing properties are ultimately governed by their magnetic states and understanding and possibly tuning them is evidently of utmost importance.
In this work we examine the \alloys alloys with Density Functional Theory simulations and $^{55}$Mn Nuclear Magnetic Resonance and combine these two methods to carefully describe their ground state magnetic order.
In addition, we compare the results obtained with the conventional generalized gradient approximation with the ones of strongly constrained and appropriately normed (SCAN) semilocal functionals for exchange and correlation. Experimental results eventually allow to discriminate between two different scenarios identified by \emph{ab initio} simulations.
\end{abstract} 


\maketitle

\section{Introduction}

Manganese based full Heusler compounds have been extensively investigated owing to the presence of multiple properties of technological interest.
These include magnetocaloric properties exploitable in energy conversion devices, such as efficient and eco-friendly magnetic refrigerators or thermomagnetic generators, giant magneto-resistance effects and shape-memory capabilities \cite{91670196,170323999,21970969}.

The fundamental physical characteristics of interest involve the structural and the magnetic transitions, their critical temperatures and the magnitude of the respective order parameters.
Controlling these phenomena by varying the chemical composition is arguably
the most straightforward method to promote industrial adoption.
This task has indeed been carried out over the years with both experimental investigations and theoretical predictions that constitute an abundant literature \cite{188452550,86005669,5881826,166886018,67736979,46645136,143525462,PhysRevB.99.224401,PhysRevB.77.012404,Kainuma2006,10.1016/0038-10987490545-6,manydft10,adma.201505571}.

In this context, Ni$_{2}$Mn$_{1+x}$Z$_{1-x}$ (with Z = Ga, In, Sn, Sb) Heusler compounds have been recently reported as an interesting class of materials \cite{21970969,CAVAZZINI201948}. They crystallize in a cubic $L2_1$ structure (austenite), composed of 4 intertwined face centered cubic lattices: two fully occupied by Ni ($8c$ Wyckoff position of the cubic cell), one by Mn ($4a$ position) and one by the non-magnetic Z element ($4b$ position) or by a random mixing of Z and Mn atoms in off-stoichiometry compounds (x $\neq$ 0) \cite{annurev-matsci-070616-123928}.
By changing the Z element and the stoichiometry they can show 
a first-order magneto-structural transition to a low-symmetry phase (martensite) well below the Curie second-order transition of the ferromagnetic (FM) austenite.
The former strongly depends on the valence electron number and can give rise to a large and tunable inverse magnetocaloric effect, a barocaloric effect and magnetic shape memory properties. Whereas, the second-order transition, ranging between 300 and 400 K, can be exploited in thermo-magnetic devices thanks to the associated reversible magnetocaloric effect.

In order to speed up the identification of other promising systems, a microscopic understanding of the compositional parameters 
governing the magnetic properties of these compounds is especially useful.
Computational approaches and high throughput material scans have already been attempted on this class of materials, but they require
accurate estimates to be available for relatively small computational efforts and are therefore generally performed only for stoichiometric samples within a mean field approximation.
A vast literature has already characterized the accuracy of Density Functional Theory (DFT) approaches for the description of the magnetic states of these
systems \cite{10.1103/physrevlett.104.176401,10.1103/physrevb.70.024427,msf.684.1,manydft1,manydft2,manydft3,manydft4,manydft5,manydft6,manydft7,manydft9,manydft11,manydft14,manydft15,manydft16,manydft17,manydft19,manydft20,manydft21,manydft22,manydft23,manydft25,manydft26,manydft27,manydft28,manydft29}.
However also in this case most of the electronic structure modeling has been conducted on stoichiometric compounds.
In addition, the investigation of the magnetic properties has been recently revisited with exchange and correlation functionals beyond the Generalized Gradient Approximation (GGA) scheme \cite{10.1103/physrevb.99.014426,38897366}, with results that partially contradict previous predictions.
More importantly, part of the new outcomes are also in contrast with experimental findings, undermining the naive expectation that, with a step forward in the Jacob ladder of exchange and correlation functionals, the description of magnetic states should accordingly improve \cite{PhysRevB.100.045126,PhysRevB.102.024407}.

In this work we assess the ground state magnetic order of the off-stoichiometry series \alloys{} and describe its microscopic arrangement combining computational methods with experimental results.
Cavazzini et al. \cite{CAVAZZINI201948} recently showed that the replacement of In with Sn induces a decrease of saturation magnetization of austenite (from 6.2 $\mu_{\mathrm{B}}$/f.u. of the In-ternary compound to 4.5 $\mu_{\mathrm{B}}$/f.u. of the Sn compound), while still preserving the FM order, and a non-monotonous variation of the Curie temperature. 
Surprisingly, this drastic change in the magnetic properties is accompanied by a very negligible lattice contraction \cite{CAVAZZINI201948}.
Preliminary DFT calculations suggested the intrinsic nature of the critical temperature variation, whereas they did not account for the origin of a substantial reduction of the macroscopic magnetic moment \cite{CAVAZZINI201948} which is still to be clarified.

Here we assess the magnetic ground state of these systems by comparing Nuclear Magnetic Resonance (NMR) measurements, that can directly probe the various Mn sites, with DFT based \emph{ab initio} predictions obtained with the ``standard'' GGA exchange and correlation energy and with the recently introduced SCAN (strongly-constrained and appropriately-normed) meta-GGA functional \cite{PhysRevLett.115.036402,10.1063_1.5094646}. A detailed electronic phase diagram is traced for the series.

\begin{figure}
    \includegraphics[width=0.7\columnwidth]{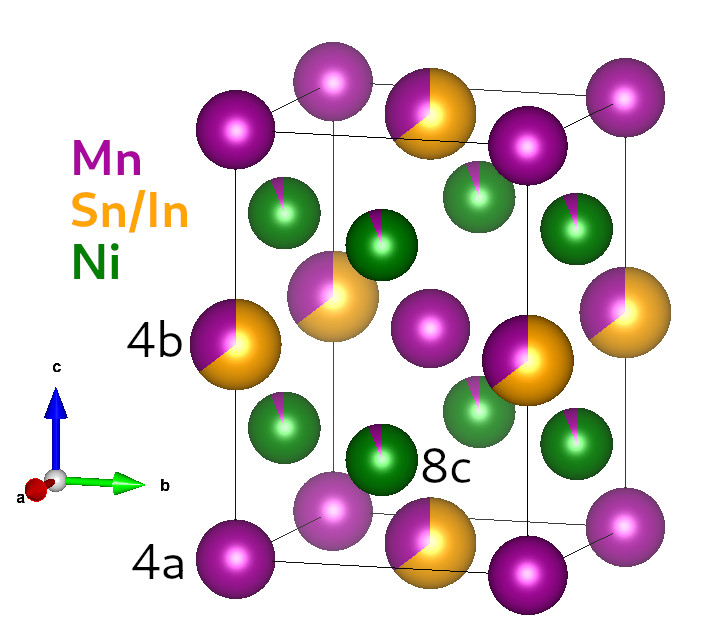}\caption{The lattice structure of \alloys in the conventional tetragonal cell. Wyckoff positions are reported according to the cubic structure.\label{fig:structure}}
\end{figure}

\begin{figure*}
    \includegraphics[width=\textwidth]{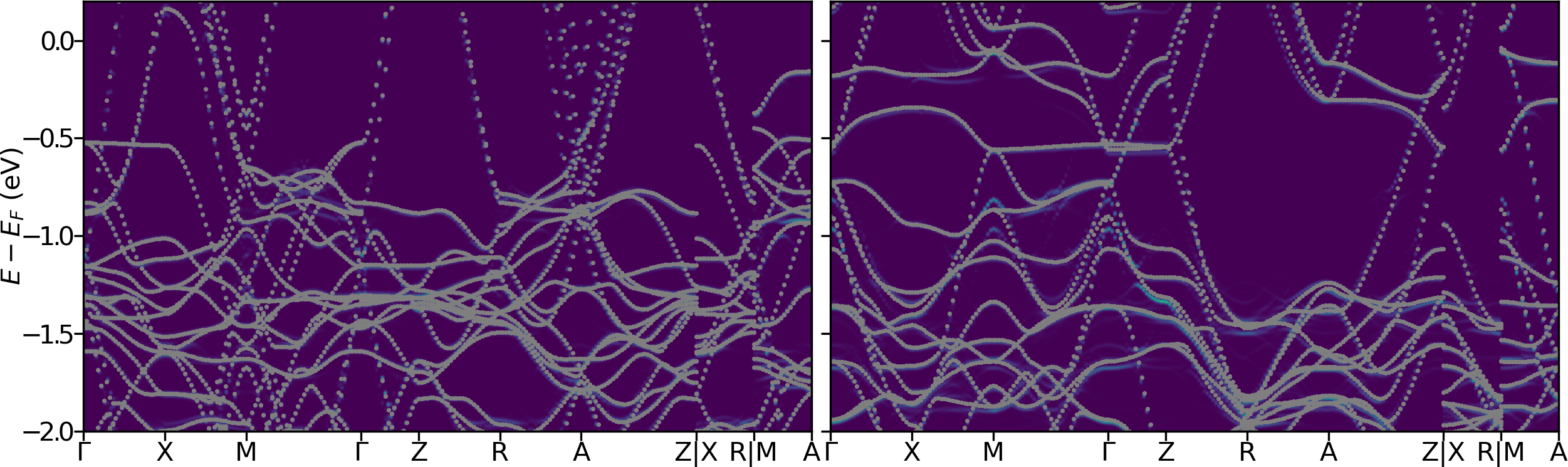}\caption{Comparison between the band structures obtained with the VCA (dotted points) and unfolded from a supercell calculation (shaded area) for \InSn with $x=0.5$. Left (right) panel are spin up (down) states. The two results perfectly match with very small deviations only in the spin minority channel.\label{fig:vca}}
\end{figure*}

The paper is organized as follow: first we present the technical details of the computational and experimental methods used in this work. We next discuss the electronic ground state of \InSn and \alloys. The predicted magnetic order is eventually used to obtain the expected low-temperature NMR spectra which are compared with the experimental measurements.

\section{Methods}

\subsection{Computational details}
DFT based calculations have been carried out with the plane-wave (PW) and pseudopotential approach using the QuantumESPRESSO \cite{doi:10.1063/5.0005082, Giannozzi2017, Giannozzi2009} suite 
and with the Linearized Augmented Plane Waves basis to perform Full Potential (FP) simulations using the Elk code \cite{elk}.
The exchange and correlation contribution was accounted for using the Perdew, Burke, and Ernzerhof (PBE) GGA functional \cite{PhysRevLett.77.3865} and the regularized strongly constrained and appropriately normed (rSCAN) functional \cite{10.1063_1.5094646} for meta-GGA.

An electronic smearing width of 0.27 eV, with the Marzari-Vanderbilt integration scheme, and a k-point grid denser than 0.3 \AA$^{-1}$ was used.
In PW simulations, when opting for PBE, we adopted ultrasoft pseudopotentials \cite{GBRV} while for rSCAN all pesudopotentials had to be replaced by the equivalent norm conserving ones \cite{ONCV} since the current implementation
of meta-GGA in {\sc Quantum ESPRESSO} lacks the support for ultrasoft pseudopotential.

     All pseudopotentials were generated with the PBE exchange and correlation functional. It was recently shown that using pseudopotentials generated with GGA in meta-GGA simulations may lead to inaccuracies \cite{1.4984939}. For this reason we validated our PW based results against FP simulations. In this latter case the meta-GGA exchange and correlation potential is obtained with an effective approximation for the functional derivative of the kinetic energy density with respect to the electron density which strongly facilitates the numerical evaluation. The details of this approach are presented in the next section, while a comparison of PW and FP results is provided in the Supplemental Material \footnote{See SM for a detailed description of the convergence of rSCAN results.}.

The plane wave cutoffs were selected in order to provide an accuracy better than 25 meV/atom for total energies and better than 4 meV/atom for energy differences (see Supplemental Material).
The resulting kinetic energy cutoffs are 612 eV and 816 eV for ultrasoft and norm conserving pseudopotentials, respectively.
These values represent the best compromise between accuracy and computational effort, the latter being a requirement to acquire enough statistics in the supercell approach described below.

FP simulations were also used to evaluate the hyperfine field at the Mn nuclei. To this aim, relevant parameters like the number of empty states, the reciprocal space integration grid and the number of local orbitals were converged against the estimate for the contact field at the Mn sites. Notably, a spin polarized solution for the Dirac equation is required to accurately account for the spin polarization of core orbitals.

In order to take into account the fractional occupations of sites $4b$ and $8c$ we used the supercell approach \cite{supercell}.
The nominal composition \alloys has been approximated with 64 atoms supercells having the slightly different distribution of atoms given by the formula Ni$_{32-\delta}$Mn$_{22+\delta}$Sn$_{10x}$In$_{10(1-x)}$.
Here $\delta$ represents the Mn content at the Ni site, while $x$ ranges from 0 to 1 and covers approximately the various concentrations of \alloys. Values of $\delta$ between 1 and 2 are the closest approximation of the real disordered structure for a given value of $x$.

The supercell simulations become tractable by performing calculations only on the symmetrically inequivalent configurations of the atomic species at the various sites.
The Sn/In compositional space has been explored using the Virtual Crystal Approximation (VCA) unless otherwise stated.

Notably, the occupation of the site $8c$ by a small fraction of Mn atoms leads to the generation of an exponentially growing number of symmetry-inequivalent atomic structures. More precisely, in the 64 atoms supercell considered here, for $x=0$ and $\delta=1$ there are 1142 symmetry inequivalent combinations, for $\delta=2$ they increase to 17286, and for $\delta=3$ they are 160882.
For this reason, a randomly chosen subset of the generated realizations has been considered in some cases, as reported later in the figures and in the main text.

\subsection{Partially deorbitalized meta-GGA}

Meta-GGA functionals depend locally on the density and kinetic energy density:
\begin{align*}
 E_{\rm xc}[\rho,\tau]=\int d^3r\,
 \epsilon_{\rm xc}^{\rm mGGA}\big(\rho({\bf r}),\tau({\bf r})\big)\rho({\bf r}),
\end{align*}
where
\begin{align}\label{tau_orb}
 \tau({\bf r})\equiv\frac{1}{2}\sum_{i=1}^N
 \big|\nabla\varphi_i({\bf r})\big|^2.
\end{align}
For its intended use, the kinetic energy density is not an independent variable
but rather an implicit functional of the
density, i.e. $\tau({\bf r})\equiv\tau[\rho]({\bf r})$.
Thus a difficulty arises when one has to determine $V_{\rm xc}$ as the functional
derivative of $E_{\rm xc}$ with respect to the density:
\begin{align}\label{vxc_mgga}
 V_{\rm xc}({\bf r})=
 \left.\frac{\delta E_{\rm xc}[\rho,\tau]}
 {\delta\rho({\bf r})}\right|_{\tau}
 +\int d^3r'\,\left.\frac{\delta E_{\rm xc}[\rho,\tau]}
 {\delta\tau({\bf r}')}\right|_{\rho}
 \frac{\delta\tau({\bf r}')}{\delta\rho({\bf r})}.
\end{align}
The last term requires the functional derivative of $\tau$ with respect to
$\rho$. This is numerically difficult to perform and requires an approach
similar to that used for the optimized effective potential (OEP)
method\cite{Yang16}. Instead, codes
typically calculate potentials derived from the derivative with respect to the
orbitals $\delta E_{\rm xc}/\delta\varphi({\bf r})$. Such an approach, however,
produces a non-local potential, which obviously violates an exact property of
the normal Kohn-Sham potential, and is referred to as {\em generalized} Kohn-Sham.
Mej\'{i}a-Rodr\'{i}guez and Trickey neatly sidestepped this problem by replacing
the $\tau$ determined from the orbitals via Eq. (\ref{tau_orb}) with one
obtained from an approximate kinetic energy density
functional \cite{MejiaRodriguez17,MejiaRodriguez18,MejiaRodriguez20}.
This so-called `deorbitalized' meta-GGA was found to produce results of
accuracy which was at least as good as the originals.
Here we adopt a similar strategy for the Elk code, but with an important
difference, namely that we use the `exact' orbital $\tau$ as {\em input} to the
functional, but use an approximate kinetic energy density functional to
compute the functional derivative in Eq. (\ref{vxc_mgga}). We term this
approach `partial deorbitalization' and find that even a fairly primitive
kinetic energy functional, like the Thomas-Fermi-von Weizsacker gradient
expansion\cite{Yang86}, yields accurate results
\footnote{Further details and validation of the partial deorbitalization method will be presented elsewhere (manuscript in preparation)}.

\subsection{Experimental details}

    \alloys{} samples were prepared following the procedure reported in Ref. \cite{CAVAZZINI201948}. Energy Dispersive Spectroscopy and X-ray powder diffraction analysis revealed chemical compositions consistent with the nominal ones within the 0.5\% and the absence of secondary phases.

The $^{55}$Mn NMR spectra were measured by means of the 
home-built phase coherent broadband spectrometer described in Ref.~\cite{allodi2005}, equipped with a helium-flow cryostat reaching a 1.4 K base temperature. Hahn spin-echoes are excited at discrete frequency points by irradiating the sample with a refocusing P-$\tau$-P Hahn radio-frequency pulse sequence, in which duration and intensities of the pulses P were optimized for maximizing the resonance signal, and the delay $\tau$ was set as short as possible (limited by the dead time of the apparatus) \cite{allodi2005}. 
Spectra are then reconstructed by taking the maximum Fourier transform amplitude of each spin-echo as the spectral density at the excitation frequency and dividing it by the frequency-dependent sensitivity \cite{Allodi_2014}.
Excited echoes are collected with a non-resonant probe circuit, comprising a small coil ($\leq$50 nH)
wound around the sample and terminated onto a 50~$\Omega$ resistor. This probe circuits allows automated frequency scans in zero and applied static magnetic field. It is worth noting that the rf enhancement of the ferromagnetic spectral components (driven by a strong hyperfine coupling) compensates the sensitivity penalty of the non-resonant probe.

\section{Results and discussion}
Let us first consider the simple \InSn composition to introduce the effect of Sn for In substitution.
In this case two Mn atoms occupy entirely the $4a$ site (in what follows \mna) while the $4b$ site has a fractional occupation shared between Mn (dubbed \mnb, 50\%), Sn and In.
Since the results for $x=1$ have already been analyzed in details in Refs.~\cite{10.1103/physrevlett.104.176401, PhysRevB.77.064417} we limit our attention to the conventional tetragonal cell (Fig.~\ref{fig:structure}) \cite{PhysRevB.101.094105}
of the austenitic phase that describes a perfectly periodic occupation of the two $4b$ sites by one Mn in the first site and by Sn or In in the second site. The fractional occupation of the $4b$ sites by In and Sn is accounted for with the VCA which is shown to be very accurate in Fig.~\ref{fig:vca} where the unfolded \footnote{The unfolding code is the same used in Ref.~\cite{5.0047266} and is available at \cite{unfoldx}.}\cite{PhysRevB.85.085201} band structure of a $2\times 2 \times 2$ 64-atom supercell with random occupation of the $4b$ site by Sn or In at $x=0.5$ concentration perfectly overlap with those obtained from the VCA at the same concentration in the conventional 8-atoms tetragonal cell.

In agreement with previous reports \cite{10.1103/physrevlett.104.176401} we found that PBE predicts a ferrimagnetic (FIM) ground state for \Sn{} with \mnb having opposite polarization with respect to \mna.
A FM ground state is instead energetically favored for \In. 
The estimated equilibrium lattice parameters and the local moments at the Mn sites, reported in Tab.~\ref{tab:firstresults}, are consistent with previous reports \cite{PhysRevB.77.064417}.

\begin{table*}
\centering
\begin{tabular}{llcccccc}
\toprule
\multirow{2}{*}{$x$} & \multirow{2}{*}{Functional (method)} & \multirow{2}{*}{Ground state} & {Lattice Parameter} & {Total Moment} &  \multicolumn{3}{c}{Moment} \\
                    &           &      &    \AA     &     $\mu_{\mathrm B}$       & \mna     & \mnb    & Ni    \\
\midrule
\multirow{3}{*}{x=1 (Sn)} & PBE (PW)    & FIM  &  5.97   & 4.1          &   3.5       &   3.7      &  0.2    \\
                          & PBE (FP)    & FIM  &  5.95   & 4.0          &   3.5       &   3.7      &  0.2    \\
                          & rSCAN (PW)  & FM   &  5.95   & 14.5          &   4.0       &   4.0      &  0.6    \\
                          & rSCAN (FP)  & FM   &  5.94   & 13.2          &  3.7       &   3.8      &  0.5    \\
\midrule
\multirow{3}{*}{x=0 (In)} & PBE (PW)    & FM   &  6.01   & 14.1          &   3.5          &  3.6    &  0.6   \\
                          & PBE (FP)    & FM   &  5.96   & 13.2          &   3.6          &  3.7          &  0.6  \\
                          & rSCAN (PW)  & FM   &  5.94   & 14.7         &   4.0       &  4.0       &  0.7  \\
                          & rSCAN (FP)  & FM   &  5.93   & 13.4         &   3.7       &  3.8       &  0.6  \\
\bottomrule
\end{tabular}\caption{Computational results for the two end-member stoichiometric compounds Ni$_{2}$Mn$_{1.5}$(Sn$_{x}$In$_{1-x}$)$_{0.5}$ with $x=0$ and $x=1$ in the austenitic phase. 
The experimental realizations of disordered Ni$_2$Mn$_{1.48}$Sn$_{0.52}$ and Ni$_2$Mn$_{1.4}$In$_{0.6}$ have lattice parameters 5.973~\AA{} and 6.017~\AA{} respectively \cite{10.1063/1.1808879}. The local moments on the various atoms are obtained by integrating the spin density in a sphere of radius 1.06 \AA.
\label{tab:firstresults}}
\end{table*}

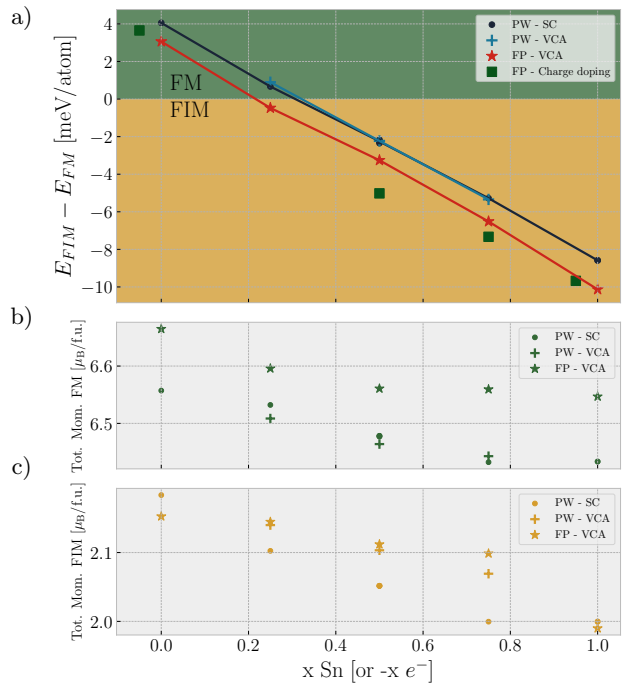
\begin{figure}[t]
\begin{tikzpicture}
    \draw (0, 0) node[inner sep=0] {\includegraphics[width=\columnwidth]{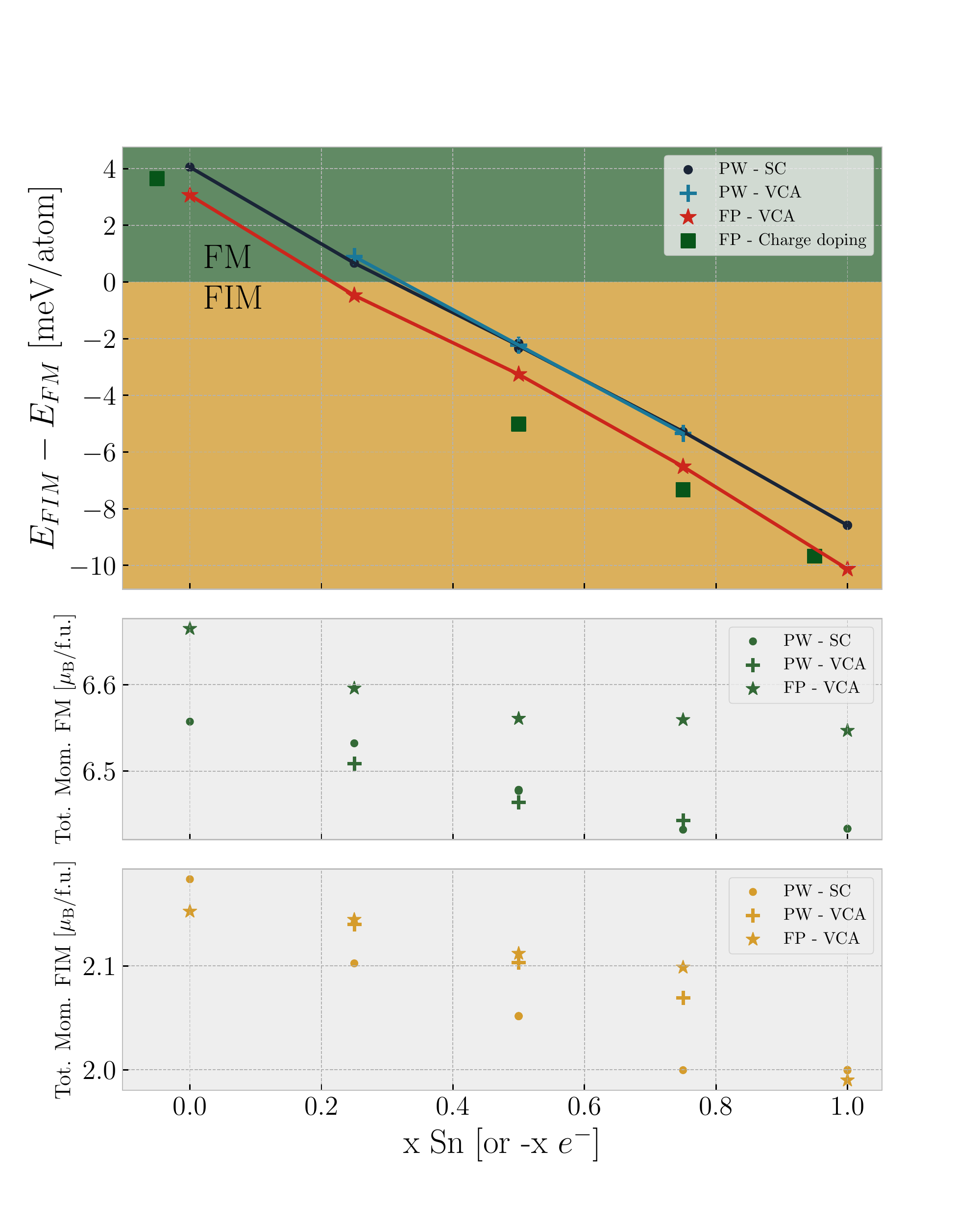}};
    \draw (-4.5, 4) node {a)};
    \draw (-4.5, 0) node {b)};
    \draw (-4.5, -2) node {c)};
\end{tikzpicture}
    \caption{Electronic properties of \InSn obtained with PBE and using different bases and approximations (see main text). In panel a) the total difference between a FM and a FIM ground state is reported for FP and PW simulations. Intermediate concentrations are simulated with VCA, with $2\times2\times2$ supercells (labelled SC) and by adding fractional electronic charges. The lines are guides to the eye. Panels b) and c) depict the magnetic moment of each composition in the FM and FIM ground states respectively\label{fig:ener}.}
\end{figure}

The substitution of In for Sn is presented in Fig.~\ref{fig:ener}, together with the trend of the magnetic moments of the FM (panel b) and FIM (panel c) states.
In the same figure it is also evidenced that the VCA reproduces very accurately also the total energy of the magnetic states of these systems, as already reported \cite{PhysRevB.77.064417, 10.1103/physrevb.70.024427}.

Independently of the magnetic order, Sn substitution suppresses slightly the total magnetization of the system, as shown in Fig.~\ref{fig:ener}b) and c).
More importantly, the additional In electron destabilizes the FIM state and eventually leads to a
magnetic phase transition at $x\sim 0.25$ according to PBE based results (Fig.~\ref{fig:ener}a).

Both PW and FP based simulations show similar trends, with the latter approach stabilizing slightly more the FIM state.
Finally, the valence electron count can also be controlled by artificially introducing electrons and a constant compensating background: the trend of the total energy difference between FM and FIM is similar to the one obtained with the previous methods but eventually departs from the real behaviour when the excess charge is close to 0.5$e^-$.

The mechanism behind the FIM to FM transition has been discussed in Ref.~\cite{10.1103/physrevb.99.014426}.
While conduction electrons contribute through a RKKY-type interaction, the magnetic properties
of these samples are dominated by local moments and nearest neighbour interactions \cite{PhysRevB.77.064417}.
Therefore, a Bethe-Slater-type behaviour \cite{10.1103/physrevb.99.014426,bs1,bs2} is to be expected and the system is found to be on the verge of the transition in light of the  inter Mn atomic distances that control the type of magnetic order.

In striking disagreement, the rSCAN functional predicts a FM ground state for both samples \cite{10.1103/physrevb.99.014426} with slightly larger absolute values for the magnetic moments, as shown in Tab.~\ref{tab:firstresults}. A very small reduction of the total magnetization, of the order of 2\%, is observed in this case.

\begin{figure*}

\begin{tikzpicture}
    \draw (0, 0) node[inner sep=0] {\includegraphics[width=\columnwidth]{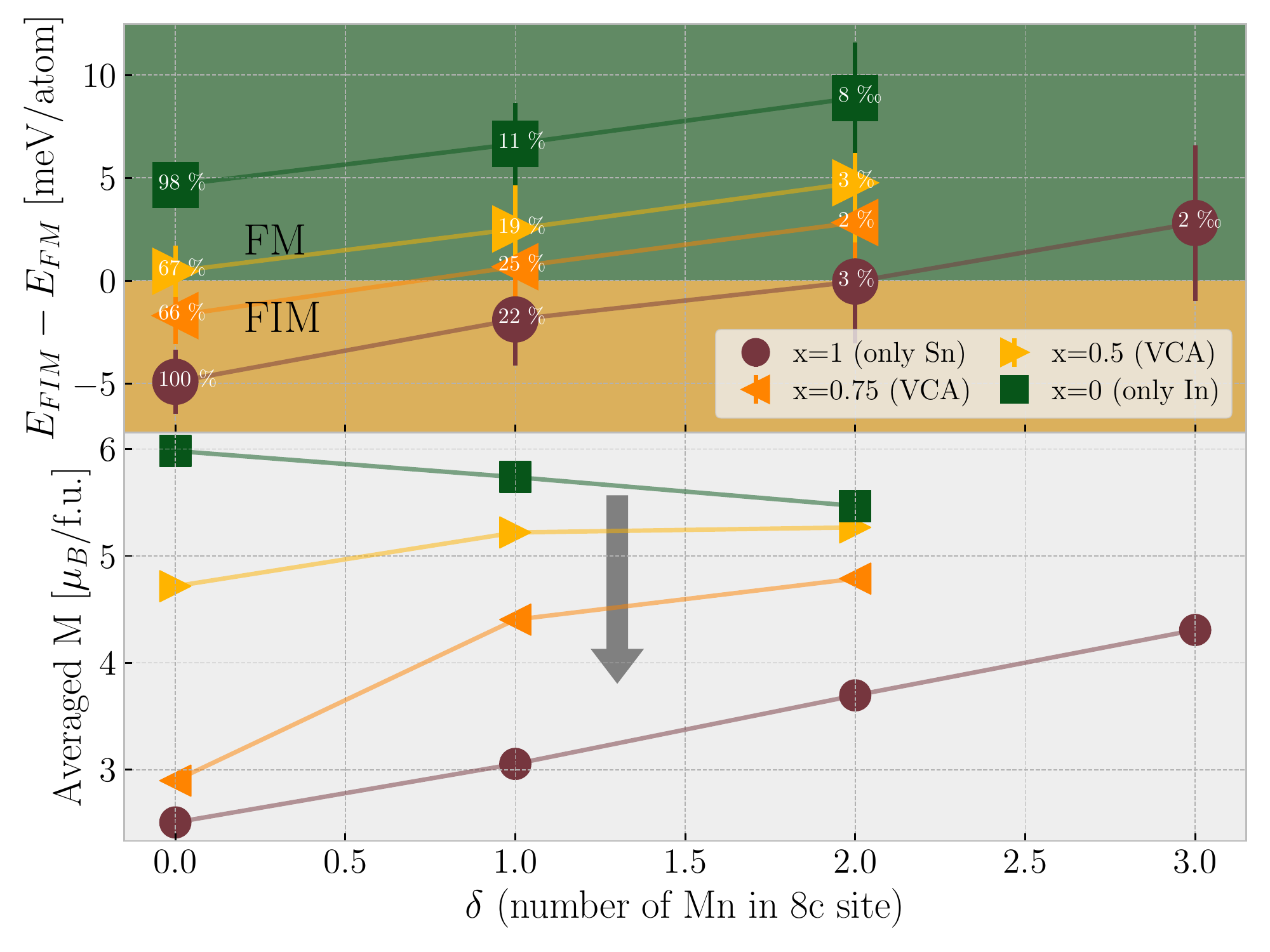}};
    \draw (8.6, 0) node[inner sep=0] {\includegraphics[width=\columnwidth]{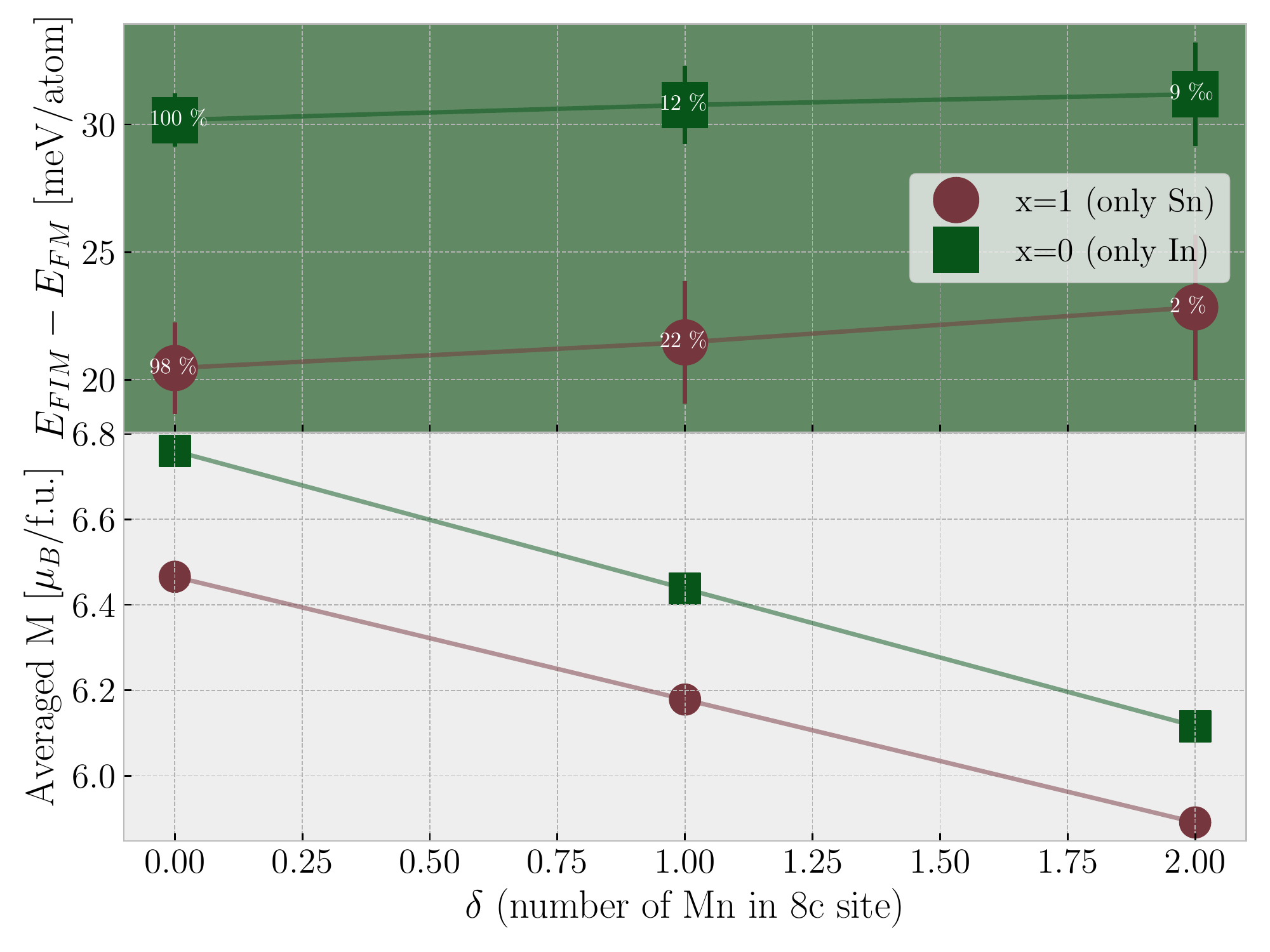}};
    \draw (-4.5, 3) node {a)};
    \draw (4.4, 3) node {b)};
\end{tikzpicture}

    \caption{Panel a): summary of relevant electronic properties for \alloys obtained  with GGA-PBE. Top graph: expected ground state magnetic order at $T=0$ obtained form the average of the various realizations in 64 atoms supercells (see text for details). The size of the symbol indicates total energy accuracy, while the error bar represents the standard deviation of the energy differences between FIM and FM states for the various realizations. The numbers inside the symbols report the percentage of symmetry inequivalent realizations that have been computed.
    The critical Mn concentration for $x=1$ is predicted to be at $\delta \sim 2$. Bottom graph: total magnetization averaged over the various supercells. The big gray arrow indicates the expected magnetic moment reduction in going for x=1 (In in 4b site) to x=0 (Sn in 4b site). Notice the  drastic reduction of the magnetization taking place between $x=0.75$ and $x=1$.
    Panel b): same as previous panel for the meta-GGA functional rSCAN.
    \label{fig:alloy}}
\end{figure*}

Let us now focus on the Ni-poor \alloys series, where two sites present fractional occupations: the $4b$ site, where both Mn, In and Sn may be present, and the $8c$ site, where a residual fraction of Ni atoms is substituted by Mn (dubbed \mnc).
The macroscopic magnetic properties of these alloys have been previously analyzed in Ref.~\cite{CAVAZZINI201948}: all samples were reported to be FM, but the evolution of the magnetization is seen to vary drastically across the compositions, with a substantial reduction of about 20\% from $x=0.75$ to $x=1$.
Both these points cannot be captured by the naive expectation that, in analogy with the results of \InSn to \alloys, the Sn for In substitution controls the magnetic ground state: PBE would predict the wrong ground state for many intermediate concentrations of the alloy, while rSCAN predicts the correct magnetic order, but the same total magnetization for $x=0$ and $x=1$, in clear disagreement with the experiment.

In light of the importance of exchange coupling interaction between localized Mn moments in these systems, we considered the effect of \mnc on the electronic phase diagram.
We explore it using the supercell approach when accounting for Mn at the $8c$ (labelled $\delta$) and $4b$ sites and with the VCA for the Sn for In substitution at the $4b$ site (labelled $x$).
Unsurprisingly, PBE and rSCAN identify rather different scenarios: the former choice for the exchange and correlation energy leads to a FIM state for $\delta=0$ and $x=1$ where \mnb are antiferromagnetically aligned with the Mn in the $4a$ site, whereas,  
on the other hand, rSCAN again predicts a FM ground state.
For $x=0$ (In end-member) and $\delta=0$, both approaches predict a FM ground state.

The presence of \mnc substantially affects the energetic balance by introducing an additional
coupling between Mn atoms in different sub-lattices where otherwise Ni mediates the indirect inter-lattice exchange interaction \cite{PhysRevB.71.014425}.
In both end-members and in all intermediate concentrations that have been tested, the spin of \mnc is anti-parallel with respect to that of \mna and has a slightly reduced moment of $m_{Mn_{c}}=2.7\mu_{B}$.
Both these results are in agreement with the qualitative expectation of the Bethe-Slater curve for the very short inter-atomic distance between Mn atoms in the $4a$, $4b$ and $8c$ sites, and show how the antiferromagnetic coupling among these sites results in the formation of a highly diluted but very strong (effective) FM coupling between \mna and \mnb.
An accurate analysis of the exchange parameters in this alloy as a function of Mn in $8c$ site and Sn/In substitution is left for future works and we instead proceed identifying the most accurate picture for the description of the magnetic ground state.

The results obtained from the various supercell simulations are averaged according the the relative weight of each symmetry-inequivalent realization and shown in Fig.~\ref{fig:alloy}, where the error bars represent the standard deviation of the total energy difference obtained as a function of $\delta$ for various values of $x$.
Surprisingly, according to PBE results shown in Fig.~\ref{fig:alloy}a, the FIM magnetic ground state of the $x=1$ end-member is strongly destabilized already at small \mnc concentrations ( $\delta\geq1)$ and a FM ground state is rapidly recovered for $\delta \geq 2$. 
In the opposite end-member, $x=0$, Mn in the $8c$ site leads to a further stabilization of the FM state.
More importantly, Fig.~\ref{fig:alloy}a shows that, for values of $\delta$ close to the experimental realization (indicated by a gray arrow), the Sn end-member ($x=1$) \emph{is at the boundary of a transition from FIM to FM} state.
At the same time, our results indicate that the competing exchange interactions are extremely composition dependent (the error bars cross the phase boundary, showing that different realization of a given concentration may favorite a FM or a FIM ground state), and, arguably, spatial inhomogeneities across the sample may lead to the coexistence of phase-separated FM and FIM states.

       It should be noted that the relaxation of atomic positions in the disordered supercells may alter the critical Mn concentration leading to the FIM to FM transition. Nonetheless, the effective FM interaction between \mna and \mnb sites mediated by the $d$ states of \mnc is arguably only slightly affected by the atomic position relaxation since the displacement of the neighbouring Mn atoms of \mnc, inspected in a limited number of realizations of Ni$_{32-\delta}$Mn$_{22+\delta}$Sn$_{10}$ with $\delta=\{1,2\}$, is found to be smaller than $\sim 0.1$~\AA. We also verified that, in the same set of simulations, the energy difference between FM and FIM states changes, on average, by 2.7 meV/atom after structural relaxation. This allows to safely establish the presence of competing magnetic states in the $x=1$ end-member.

The results obtained with the rSCAN, presented in Fig.~\ref{fig:alloy}b, are again strikingly different. While the magnetic moment on \mnc is still found to be directed opposite to \mna, the magnetic ground state of all the realizations that have been considered is FM. It is therefore to be expected that the addition of Mn on the Ni site does not alter the stability of the ground state which is instead more substantially affected by the valence electron count. The only relevant change observed for $\delta \geq 1$ is a small reduction of the total magnetic moment, that, incidentally, is also present in PBE results and trivially originates from \mnc.

Summarizing, the overall pictures that can be traced with PBE and rSCAN are very different: our set of samples appears to be on the verge of a phase transition according to PBE, while rSCAN results describe a different scenario where all compounds are ferromagnetic irrespective of the concentration of \mnc and valence electron count.

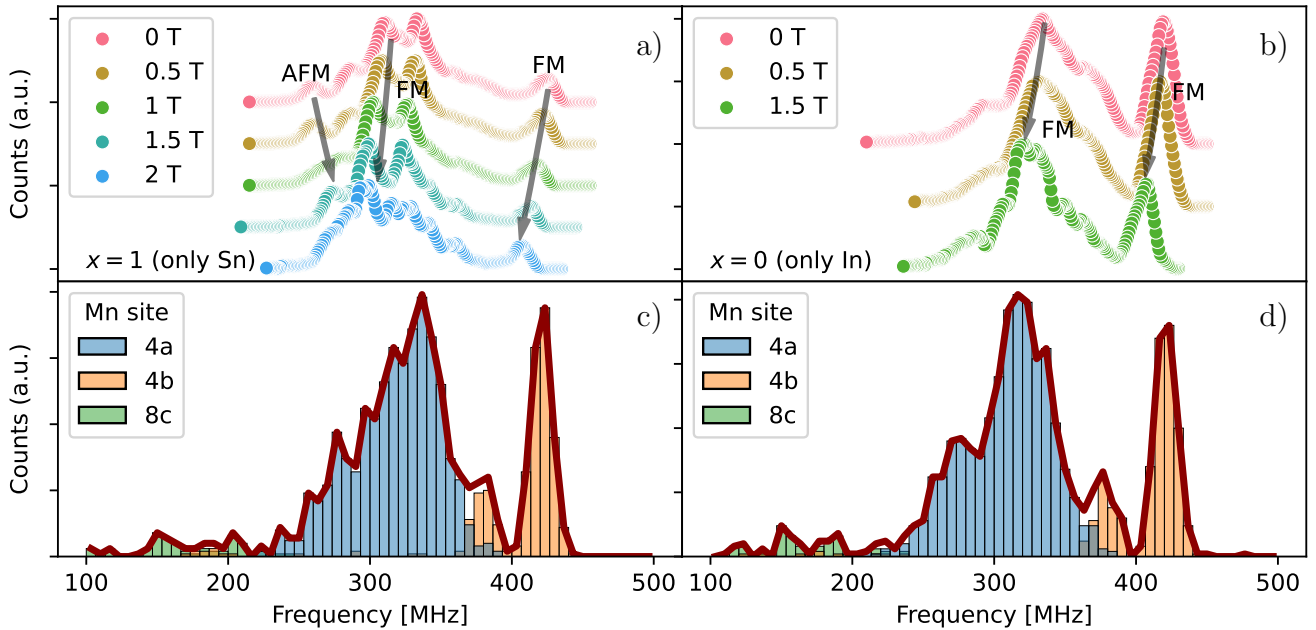
\begin{figure*}[t]
\begin{tikzpicture}
    \draw (0, 0.1) node[inner sep=0] {\includegraphics[width=\textwidth]{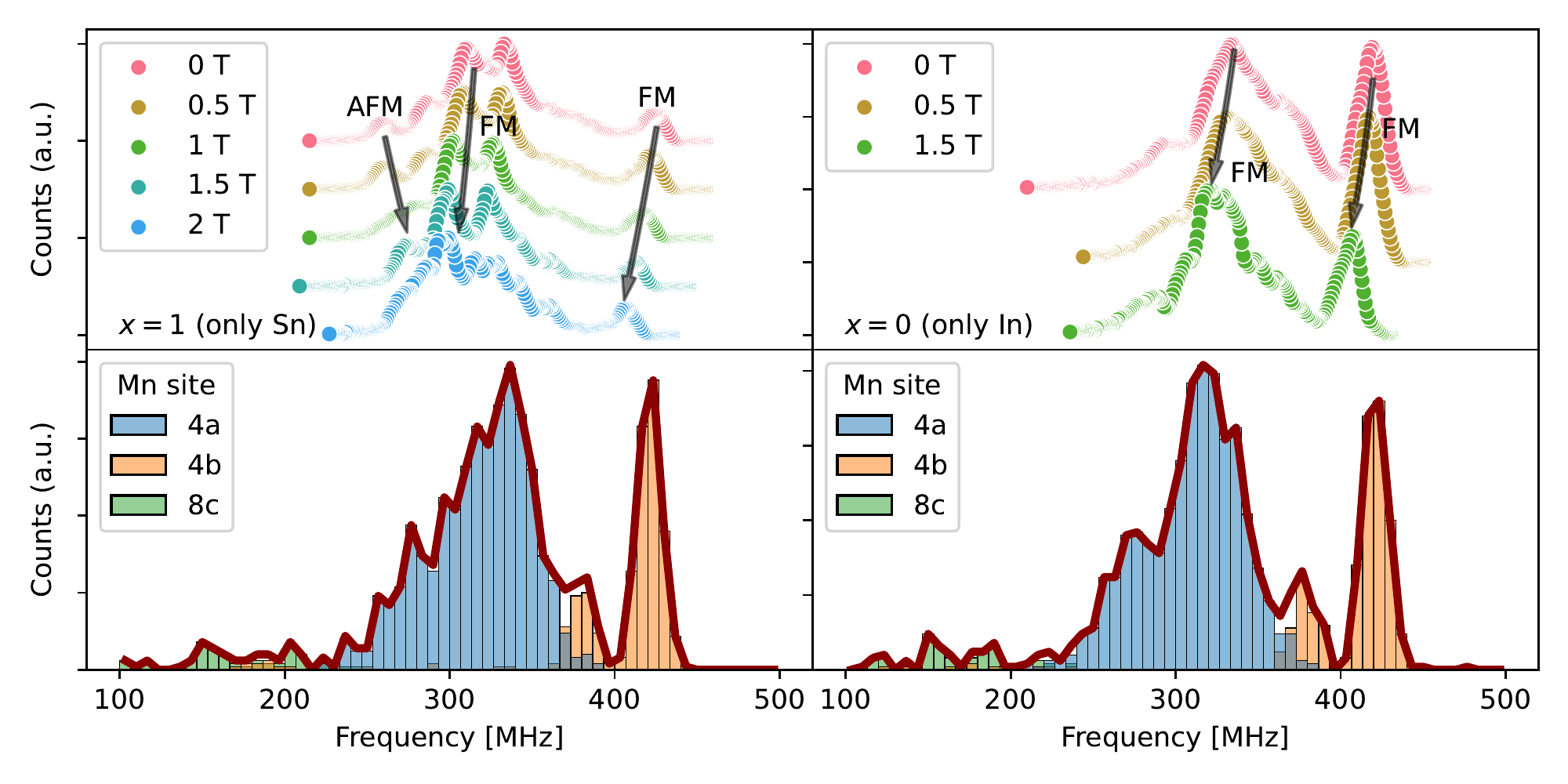}};
    \draw (-0.1, 3.7) node {\large a)};
    \draw (8.2, 3.7) node {\large b)};
    \draw (-0.1, 0.1) node {\large c)};
    \draw (8.2, 0.1) node {\large d)};
\end{tikzpicture}

\caption{Comparison of experimental results (panels a and b) and computational estimates (panels c and d) for the $^{55}$Mn NMR spectra. The gray arrows superimposed to the NMR data obtained as a function of applied field show the trend typical of ferromagnetic and antiferromagnetic coupling of different portions of the spectra. The predicted contact field at Mn sites is reported separately for \mna, \mnb and \mnc while the thick red line is the sum over all contributions from the various Mn sites. \label{fig:nmr}}
\end{figure*}

At first sight the second option seems to better describe the experimental evidence of a FM ground states for $T\to 0$~K across the whole series, but the reduction of the total magnetization of about 70\% observed as a function of In for Sn substitution is instead accounted for by PBE results provided that, for $x=1$, both FM and FIM states coexist in the sample (Fig.~\ref{fig:alloy}a, lower panel).

In order to verify which picture more closely describes the magnetic order of \alloys, we conducted $^{55}$Mn nuclear magnetic resonance (NMR) experiments at 1.4~K and computed the hyperfine field at the Mn sites with a FP approach.
NMR is the ideal tool to verify which magnetic state is actually realized in these compounds since, through the hyperfine coupling of $^{55}$Mn nuclei with surrounding electrons, it allows to probe the whole sample without introducing spatial averages.

The spectra obtained for the ternary \alloysn and \alloyin compositions are reported, together with the \emph{ab initio} predictions, in Fig.~\ref{fig:nmr}.
The complex structure of these spectra is the consequence of the compositional disorder present in both the $x=0$ sample (Fig.~\ref{fig:nmr}a) and the $x=1$ sample (Fig.~\ref{fig:nmr}b). Still the trend of the various peaks as a function of applied filed allows to distinguish two separate behaviours: the vast majority of the contributions shift to lower fields as the external applied field is increased, typical for ferromagnetic coupling, while a smaller feature at low frequency in the $x=1$ sample (Fig.~\ref{fig:nmr}a) clearly shows the opposite trend, shifting to higher frequencies, as shown by the gray arrow, indicating the presence of antiferromagnetic coupling.

Our full potential simulations show that the hyperfine field is dominated by the contact term, with orbital and pseudo-dipolar contributions being of the order of 1 T \footnote{The evaluation of orbital and pseudo-dipolar term requires non-collinear simulations that are prohibitively expensive. The values quoted in the main text have been obtained with two realizations of the full Sn and full In compounds.}. The distortion of the lattice in the various realization slightly affects the hyperfine field at Mn sites which in this case changes by less than 1 T for the cases that we have tested.
Since accounting for structural relaxations and/or orbital and pseudo-dipolar contributions would make the simulations computationally  unfeasible, we limit our analysis to the contact interaction assuming an uncertainty of the order of a few Tesla.

The computational predictions for the NMR spectra were collected on a set of 121 realizations of the $\delta=1$ and $x={0,1}$ supercells, totaling 2783 Mn contact hyperfine couplings obtained in the FM ground state\footnote{We label this state FM although the moment on \mnc is antiparallel to that of all other manganese atoms in the system.}. In this case the contact term at the \mna and \mnb sites has the same sign, with the latter larger than the former and of the order of 30 T. 
As expected, our simulations show that the hyperfine field is found to be directed anti-parallel to the $d$-orbital spin as it mainly originates from the core-orbitals spin polarization in transition metal ions.
A systematic underestimation of the hyperfine coupling of about 7.5 T is identified by the comparison of the distribution of contact fields obtained from first principles with the experimental ones.
With a constant shift of the aforementioned amount, the expected frequency distribution obtained \emph{ab initio} is shown in Fig.~\ref{fig:nmr}c) and d).
This allows to unambiguously assign the high frequency peaks at $\sim 450$~MHz to \mnb while \mna atoms contribute to the broader distribution centered at $\sim 300$~MHz.
The smaller, antiferromagnetic coupling of \mnc, predicted at very low frequency (see Fig.~\ref{fig:nmr}c and d), is instead apparently not observed (its resonance peak is probably hidden underneath the low-frequency tail of the broader \mna resonance).
The presence of antiferromagnetically coupled Mn must therefore be attributed to the presence of FIM phases in the sample where the signal produced by \mnb is expected to give rise to the peak at about 260 MHz.

This conclusion is further supported by a quantitative analysis of the NMR data. Experimentally, only the Sn-rich samples are characterized by the clear presence of a fraction of nuclei sensing a positive hyperfine field and therefore coupled to counter-aligned spins in a ferrimagnetic order. This fraction of antiferromagnetically aligned Mn atoms is much larger than the one expected from the \mnc sites alone and must be attributed also to \mnb spins that are opposed to the majority \mna ones. Actually, the fact that an antiferromagnetic response is only detectable for $x = 1$, seemingly rules out that such signal may originate at all from \mnc spins, whose antiferromagnetic order is independent of composition. This indicates that \alloys{} for $x \to 1$ is characterized by the presence of both FM and FIM domains, whose ordered \mnb spin populations are quantified are quantified in the 8.8(7)\% and 6.1(7)\% of the number of moments on the \mna site, respectively. By assuming the occupation of Mn nuclei to be 1 on 4a site and 1/3 on the 4b site, we can conclude that the 26(1)\% of \mnb is FM ordered, the AFM fraction represents the 18(1)\% of the \mnb moments, while the rest of \mnb is magnetically disordered at 1.4~K.

\section{Conclusions}
The phase diagram of off-stoichimetric \alloys has been characterized by a careful interpretation of NMR results based on DFT simulations.
We have shown that the ground state of \alloys is controlled by both valence electron count and, more importantly, by substitutional Mn in the $8c$ site.
The residual occupancy of the Ni site by Mn defines the magnetic ground state and we further show that the $x=1$ member of \alloys is at the critical concentration for the transition between a FIM and a FM state. 
This explains the unexpected reduction of the magnetic moment of the series for $x\to 1$ at almost constant unit cell volume and FM transition temperature.
Future research to extend this study should also examine the structural stability of the alloys as a function of their stoichiometry and the effect of lattice strain caused by the replacement of larger In/Sn atom with smaller Mn atom \cite{Nevgi_2020}.
These results may also be of interest for the design of new materials with giant tunable exchange biases \cite{ma2021}.

Finally, an \emph{ab initio} description of magnetism in these systems is more accurately provided by the PBE GGA functional for the exchange and correlation energy.
Indeed, the predictions obtained with rSCAN overestimate the stability of the FM state and therefore miss the possibility of a competition between different spin alignments for the Mn atoms in the $4b$ sublattice which is instead confirmed by our experimental findings. These results cast further doubt on the accuracy of rSCAN for the description of localized magnetic states.

\section{Acknowledgements}
The authors wish to thank Pietro Delugas for fruitful scientific 
discussions.
The computational work of P.B. was supported by the SUPER, Action 1.5.1, 
POR-FESR 2014-2020 Emilia Romagna project and by CINECA IsB20\_PRISM and IsC71\_DMSPER grants.
P.B. dedicates this work to his fellow citizen Patrick George Zaki and to all political prisoners in Egypt.
The first three authors contributed equally to this work. 
\bibliographystyle{apsrev4-1}
\bibliography{bib}

\clearpage

\section*{Supplemental Information for Magnetic phase diagram of the austenitic Mn-rich Ni-Mn-(In,Sn) Heusler alloys}

\subsection{Convergence and validation of rSCAN results}

The convergence and the validation of our supercell results using regularized strongly constrained and appropriately normed, rSCAN \cite{10.1063_1.5094646}, meta-GGA exchange-correlation approximation require special attention.
Firstly, some properties, including magnetic moments, may convergence slowly with respect to the cutoff energy of the plane wave expansion. Secondly, and more importantly, while many studies use pseudopotentials generated with the GGA to perform meta-GGA simulations, it has recently been shown that this may lead to inaccuracies \cite{1.4984939}.

\begin{figure*}
    \centering
    \includegraphics{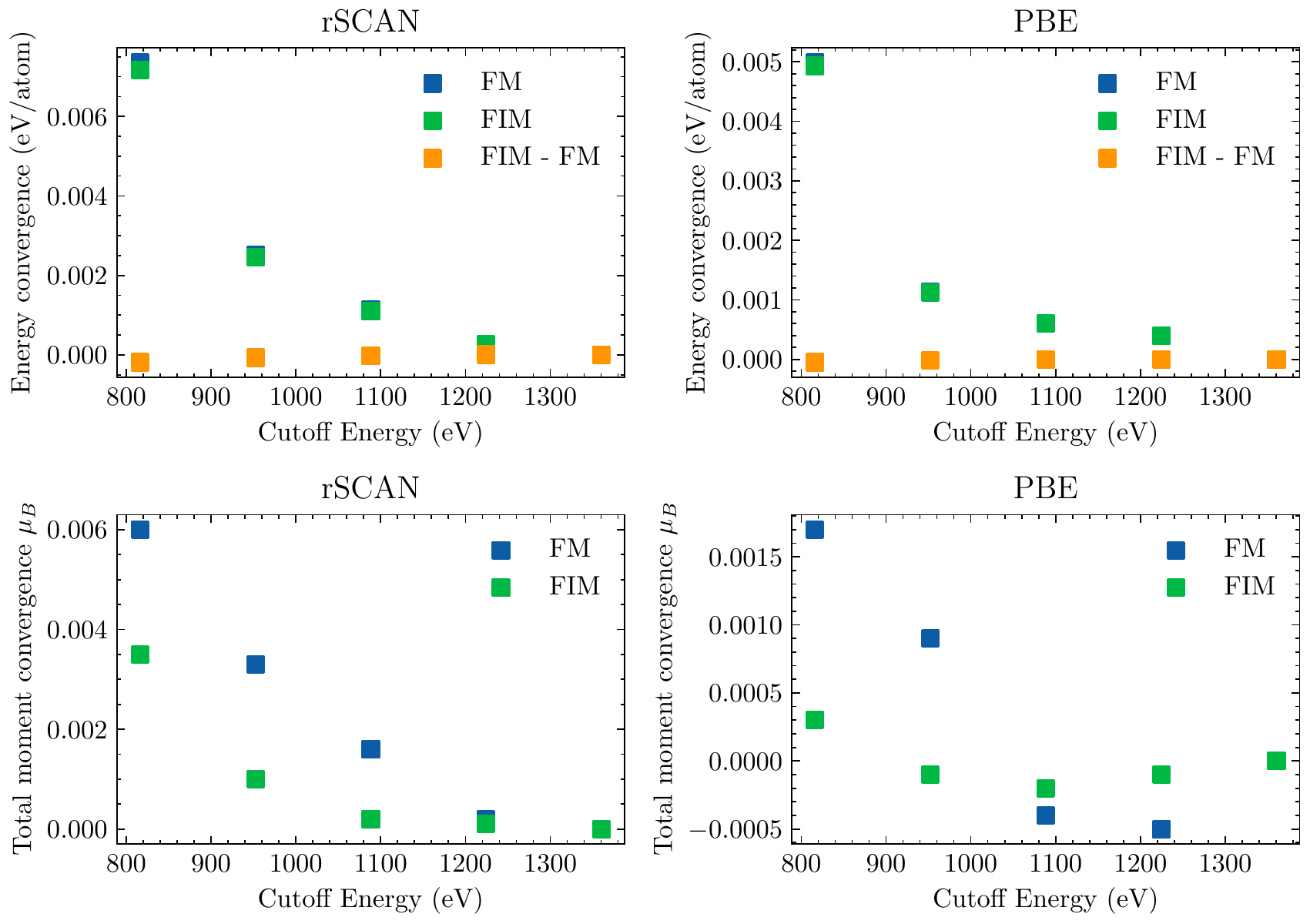}
    \caption{Convergence of total energy and total moment against plane wave energy cutoff for PBE and rSCAN simulations in \Sn. In both cases norm conserving pseudopotentials were used. The plots report the difference between the values obtained with the largest basis size (last point on the right in each plot) and those obtained with smaller cutoffs.}
    \label{fig:convergenceSn}
\end{figure*}

\begin{figure*}
    \centering
    \includegraphics{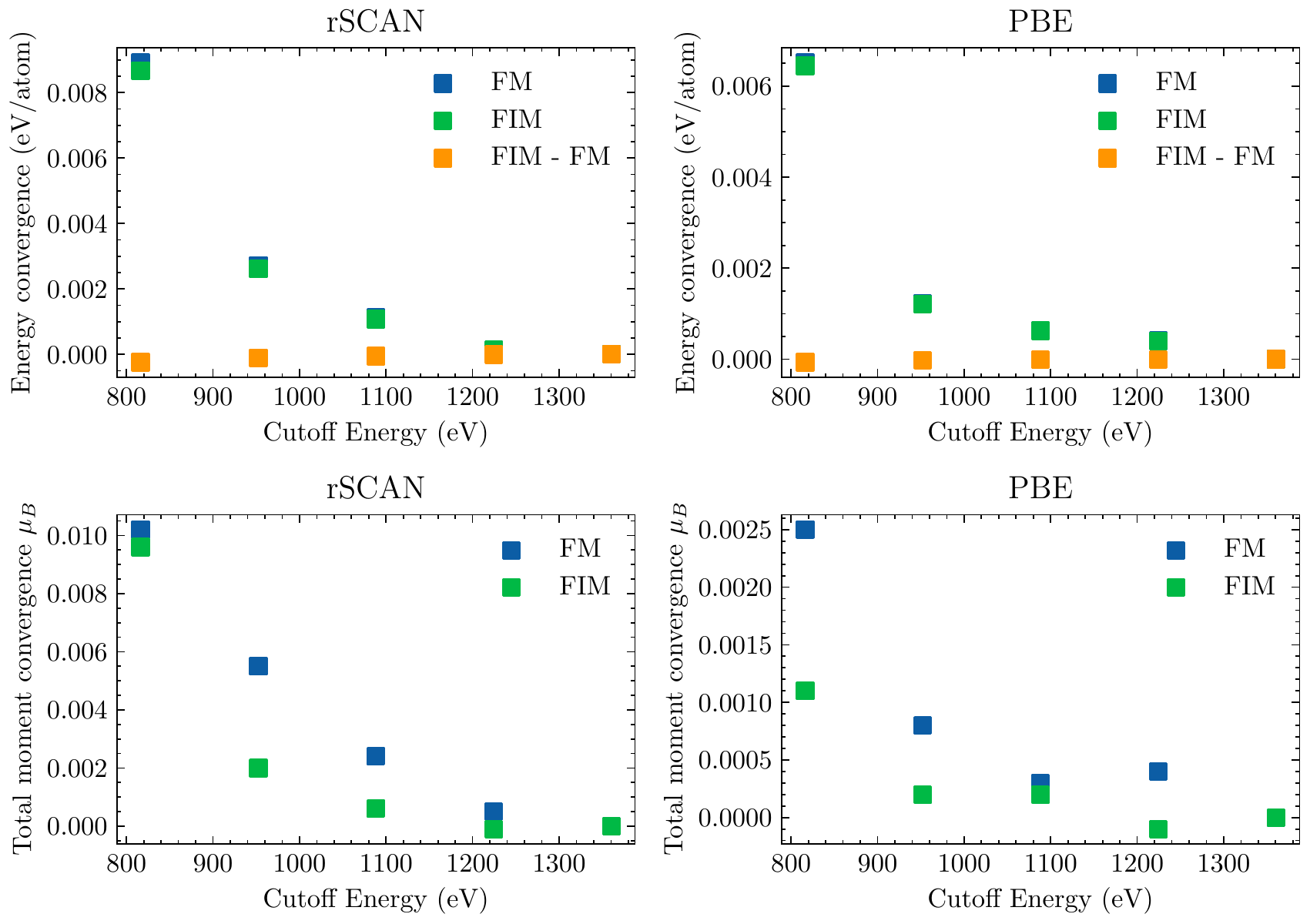}
    \caption{Convergence of total energy and total moment against plane wave energy cutoff for PBE and rSCAN simulations in \In. In both cases norm conserving pseudopotentials were used. The plots report the difference between the values obtained with the largest basis size (last point on the right in each plot) and those obtained with smaller cutoffs.}
    \label{fig:convergenceIn}
\end{figure*}

For this reason we present here the convergence tests for our plane wave (PW) based results and a comparison between converged results obtained with PW and full potential (FP) LAPW based simulations performed with the code Elk on \In and \Sn.

\begin{figure*}
    \centering
    \includegraphics{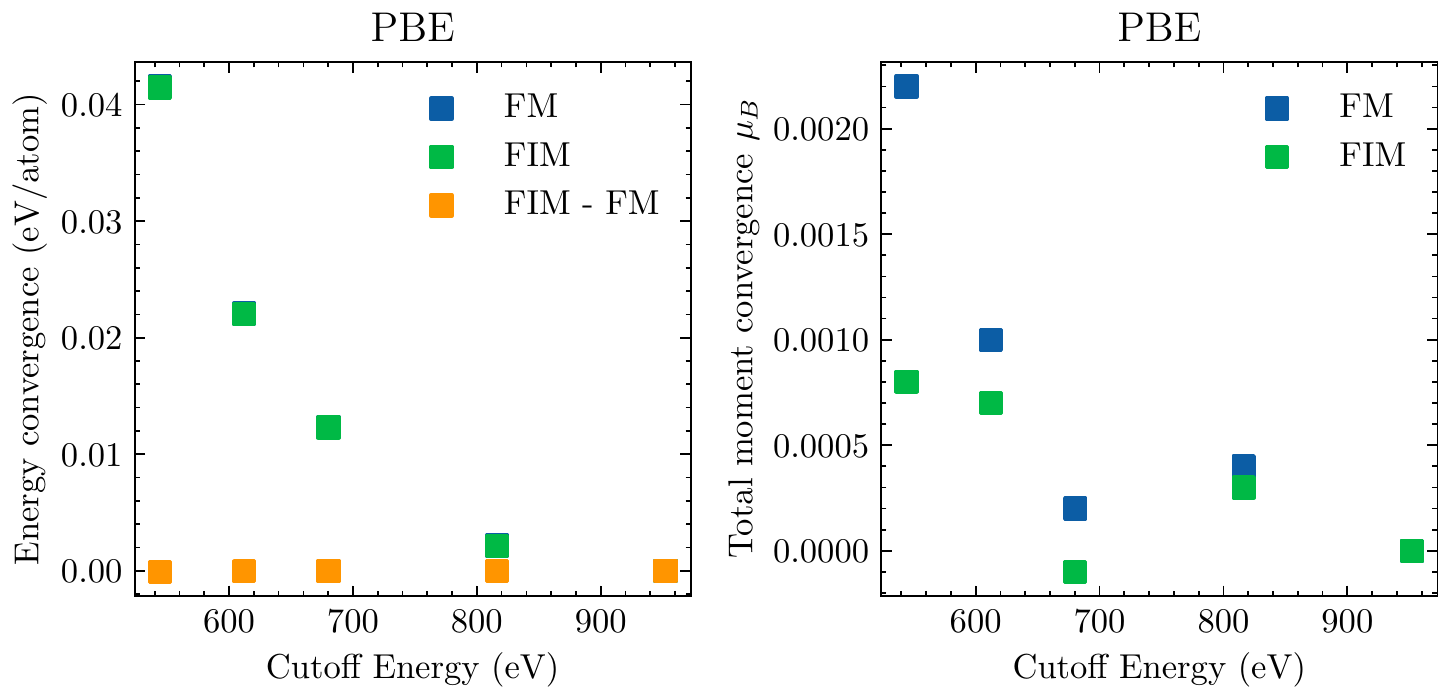}
    \caption{Convergence of total energy and total moment against plane wave wavefunction cutoff for PBE and ultrasoft pseudopotentials in \Sn. The plots report the difference between the values obtained with the largest basis size (last point on the right in each plot) and those obtained with smaller cutoffs.}
    \label{fig:convergenceUSPP}
\end{figure*}

Fig.~\ref{fig:convergenceSn} and Fig.~\ref{fig:convergenceIn} report the convergence of the total energy and the total moment of both the FM and the FIM states (see main text) in \Sn and \In as a function of the cutoff energy used in plane wave expansion and using norm conserving pseudopotentials. All values are reported with respect to the reference value obtained at the maximum cutoff energy of 1361 eV.
For completeness we also report the convergence tests performed with ultrasoft pseudopotentials on \Sn in Fig.~\ref{fig:convergenceUSPP}.

The convergence with respect to the reciprocal space sampling does not pose significant problems and the selected reciprocal space grid density (see main text) guarantees that the accuracy is only limited by the dimension of the basis set.

\begin{figure*}
    \centering
    \includegraphics{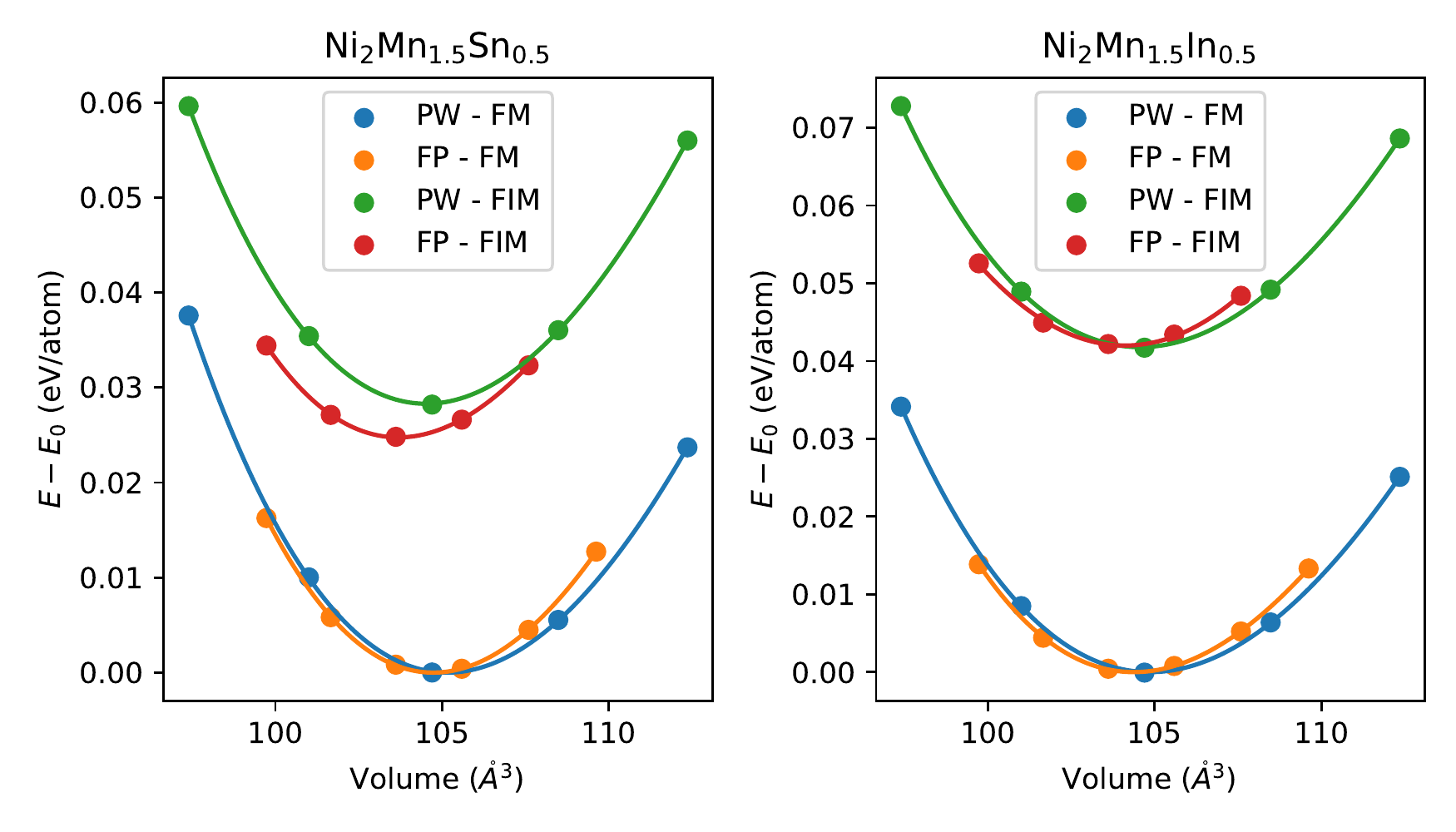}
    \caption{Comparison of FP and PW based results obtained with rSCAN for \InSn with $x=0$ and $x=1$. The lines are fit to the stabilized jellium equation of state.}
    \label{fig:fp}
\end{figure*}

In order to address the second point, i.e. the accuracy of PW based rSCAN simulations, we report in Fig.~\ref{fig:fp} the total energy of \Sn and \In as a function of the unit cell volume for both FM and FIM states obtained with the PW based code and the partially deorbitalized implementation of meta-GGA recently introduced in Elk and described in the main text.

In both FP and PW results, the total energy of the FM state at the equilibrium volume,
obtained with a fit to the stabilized jellium equation of state \cite{PhysRevB.63.224115},
has been subtracted.
As shown in Fig.~\ref{fig:fp}, the FM order is the ground state for both \In and \Sn when using rSCAN and
the predicted equilibrium volume obtained with both approaches matches very well.
The FIM states have higher energies by about 30 and 40 meV/atom for \Sn and \In respectively.
These energy differences are well reproduced by the PW simulations and the equilibrium volume of this higher
energy state is only slightly deviating from the one obtained by FP simulations in \Sn.

These results validate the PW based supercell simulations of \alloys{} discussed in the main text.


\end{document}